# Mid-infrared photodetector based on 2D material metamaterial with negative index properties for a wide range of angles near vertical illumination

E. Ponizovskaya Devine[1*]

[1]University of California, Davis, CA 95616

**Abstract:**

We present the vertically illuminated Si compatible mid-infrared photodetectors based on graphene that forms negative-index metamaterial. The coupling into a 2D layer at the angles near normal to the surface is negligible without the help of a metamaterial. We propose a low loss metamaterial, the dielectric structure with the nano/microholes through 2D materials that supports a resonance and enhances quantum efficiency in a wide range of incident angles.

*Keywords*: mid-infrared, 2D materials, photodetectors

[*1]eponizovskayadevine@ucdavis.edu

## 1. Introduction

Recently, 2D materials attract attention for multiple applications including photodetector in infrared and mid-infrared [1]. The high mobility of carriers in graphene and other 2D materials makes it a preferable choice for fast device performance. At the same time, embedded into Si 2D materials are able to overcome the challenge that Si-compatible designs in infrared are facing the use of III-V and II-VI compound semiconductor materials. Graphene is a typical 2D material with a wide range of applications. One of them is enhanced X-ray photon response [2], AgInZnS-graphene oxide (GO) nanocomposites and their application in-vivo bioimaging [3] enhanced the performance of light-controlled conductive switching in hybrid cuprous oxide/reduced graphene oxide (Cu2O/rGO) nanocomposites [4]. This study is focused on mid-infrared applications. Graphene and other 2D materials like MoS2 or p-AsP [5-6] embedded in Si could be the solution to the problem.

However, absorption of 2D materials for the vertical illumination in mid-infrared is negligibly small and there is a demand for metamaterials that facilitate the confinement of the illumination is a one-atom layer thickness. The studies [7] used metal particles that help to couple the plasmonic excitation at the surface into graphene. The disadvantage of the approach is significant losses in the metal particles. The paper [8] have shown a strong but very angle sensitive plasmonic coupling in the graphene sheet and in an array of the graphene sheets. The negative index materials exhibit the superlensing properties and, thus, enable good infrared confinement [9-10]. Recent research discussed the theoretical possibility to use negative index materials helps confine the plasmons in graphene [11]. The materials with the negative index properties could provide very small light confinement due to the superlens effect [12]. The negative refractive index (NRI) materials based on metallic structures have significant losses. In the manuscript, we study the possibility to design the metamaterial that exhibits NRI due to the photocurrent resonance with the device topology. As a result, the metamaterial, which consists of graphene layers and the dielectric, produces the magnetic resonance in a similar way as the fishnet [9], and the effective refraction index is negative. The normal incident infrared light in negative-index metamaterial confines at 2D material sheets. Thus, there are no losses related to metallic parts in plasmonic partials or split-ring resonators, instead, the photocurrent excited in 2D materials produce the desired effect.

Meanwhile, the applications using light-trapping structures have shown to be effective for enhancing absorption and expanding the operating band into longer wavelengths [9]. The micro-holes structures were used to scatter the normally incident light into the lateral propagation, forming guided modes, and increasing that allows using the thin layers of absorbing material more effective [12-14]. The effect was studied for mid-infrared material to increase the wavelength range of the conventional infrared materials [15]. Our study is focused on the dielectric structures that could provide good confinement and avoid losses of the metal structures. For this purpose, we consider layers of graphene in Si, which is transparent in mid-infrared, and a microhole array that facilitates to couple a vertical illumination into a plasmon in the graphene layer. Without nano- or micro-structure, the coupling could happen only at a certain angle. We have shown that with the designed metamaterial we can achieve a coupling in a wide range of angles including the normal incidence. The infrared radiation induces the current in the graphene producing the resonance effect that could support the backward propagation and negative index properties [10].

The holes in the layer of conductive material form the fishnet structure that supports magnetic resonance in a wide range of angles near vertical and provides negative index properties. We can use magnetic resonance to enhance the lateral propagation of the plasmons in 2D material arranged in a structure similar to fishnet. Instead of using a graphene sheet between the negative index material to focus the illumination on graphene as it was proposed in [13], we use the NRI properties of the graphene in Si with micro holes. Besides that, we can connect the 2D materials to the electrodes in the photodetector application.

## 2. Design and Simulations

We solve the Maxwell equations for the vertically incident illumination of the structure presented in Fig.1. The numerical simulations were done for a microstructure that consisted of two layers of graphene embedded into Si on SOI substrate with light trapping nano or micro-holes as shown in Fig.1. The graphene layers (blue lines in Fig.1c) were positioned with a gap between them and connect to the electrodes (yellow areas in Fig.1a) in *xy* plane.

The holes that etched through the graphene and the Si to the substrate. The micro-hole structure resonates with the infrared radiation and produces the modes in the direction parallel to the substrate. The metamaterials are formed by the graphene layer and the holes in dielectric support modes that could be confined near the location of the graphene layers. The micro-holes in the structure are chosen to be rectangular holes in a square lattice with a size of 0.7x1.5 microns and a period of 2.5 microns and a depth of 0.75 microns.

Graphene is one atom thick honeycomb lattice of sp2 bonded carbon atoms optical conductivity was studied in [16,17]. The optical response is characterized by its surface conductivity related to its Fermi energy. As we know for a monolayer of graphene the energy bands near K point in the 2D Brillouin zone are cones $e_{1,2}(k) = \pm uk$, where $u$ is a constant velocity parameter $10^8$ cm/s. The graphene conductivity in the plane of the material is [16]:

$$\sigma(\omega) = \frac{\sigma_0}{2}\left(\tanh\frac{h\nu+2E_F}{4k_BT} + \tanh\frac{h\nu-2E_F}{4k_BT}\right)$$
$$-i\frac{\sigma_0}{2}\log\left(\frac{(h\nu+2E_F)^2}{(h\nu+2E_F)^2+(2k_BT)^2}\right) + i\frac{\sigma_0}{2\pi}\frac{4E_F}{h\nu+ih\gamma}$$
$$\sigma_0 = e^2/(4h).$$

(1)

Here $E_F$ is the Fermi energy relative to the Dirac point, $e$ is the electron charge, $\nu$ is the frequency, $h$ is the Planck's constant, $k_B$ is the Boltzmann constant, $T$ is the temperature, and $\gamma$ is the intraband scattering rate. The graphene model takes into account interband transitions and the Drude-like intraband conductivity. The conductivity with E-field perpendicular to the $xy$ plane is 0. The Finite Difference Time Domain (FDTD) method with optical conductivity for graphene given by (1) and the Bloch boundaries around the unit cell and absorbing boundaries perpendicular to the illumination were used for the simulation. The p-polarization with E-field in the plane xy-plane was chosen. The coupling into a plasmon monolayer was studied in [17]. If the intraband contribution is dominant the dispersion relation for the surface plasmon in the monolayer of graphene [8] will be:

$$k_{SPP} = k_0\sqrt{\varepsilon_d + \left(\frac{2}{k_0\xi}\right)^2} \quad k_0 = \frac{2\pi}{\lambda} \quad \xi = \frac{\eta_0\sigma}{i\varepsilon_d k_0}$$

(2)

$\eta_0$ is the impedance of air.

From the boundary conditions for the electric field $E$ and magnetic field $H$ at the wall of the holes, the constrain on the $k$ should be found. As a result, the radiation is spread with a non-zero lateral $k$.

The coupling into graphene must satisfy the boundary conditions that assume the constraints for non-zero $k$ in the $xy$-plane. In the case of the holes, the infrared light is dispersed in the $xy$ plane as it was shown in [14]. However, without the resonance, the enhancement will be weaker and will depend on modal density that corresponds to the plasmon constraints. The resonance enhances the absorption even stronger. The interaction of plasmons in the multiple layers would shift the resonance. To study the resonance effects the FDTD method was used.

## 3. Results and Discussion

The illustration for the plasmon condiments in the metastructure is presented in Fig.2. As we can see even the normal incident illumination resonates with the layers of 2D material similarly to fishnet nanostructures [8] and produces plasmon in 2D layers that reveal themselves in the Poynting vector distribution as a hot spot of the energy. The source is a plane wave with a wavelength range between 5 and 10 microns. The Poynting vector distribution is shown for normal incidence (Fig.2 top) and an angle of 5 degrees with normal. The Poynting vector amplitude is normalized to the incident illumination power.

The plasmon is exited in graphene for both cases. We studied several configurations for the positions of the 2D layers. The first design has one 2D layer in the middle of the 1.5 microns Si layer. The second one contains two layers with a distance between them of 0.56 microns.

The absorption, reflection, and transmission for two graphene layers separated by 0.56 microns, embedded in 1.5 microns of Si transparent at mid-infrared are shown in Fig.3. The resonance behavior provides high absorption. Elliptical holes or rectangular holes are polarization-dependent. Thus, the micro-hole structure can be tuned to detect the polarization-dependent illumination. The absorption without holes is negligibly small and its value is a fraction of a percent. The micro-holes increase the absorption due to redirecting the radiation parallel to the graphene layer and coupling it into a resonant plasmon. The resonance can be tuned for a different wavelength range by changing the distance between the graphene layers.

The eigenvalues of the system determine the structure resonance. As one can see from Fig.3 the structures have multiple that corresponds to the peaks in absorption. Some of them, such as the peak near 7.5 microns, are extremely narrow for wide frequency range photodetectors. The compromise between the efficiency and range could be achieved near 8.3 microns. The plasmonic resonance depends on the distance between the graphene layers and the simulations show that the position of the resonance changes with the number of layers and the distance between them (Fig.3). The typical resonance transmittance and reflectance behavior are seen in Fig.4a. The size of the openings, the width, the distance between graphene layers was optimized for the range between 8 and 9 microns. However, the narrow resonance at 7.3 microns could be useful in the systems that need signal filtration from the other frequencies. The research is focused on the photodetectors vertically illuminated in a wide range of angles. The simulations (Fig.4b) shows that the absorption peak remains the same for a wide range of angles. The simulations were made in a wide range of incident angles up to 15 degrees near vertical. The peak around 8.3 microns changes insignificantly.

**4. Conclusions**

Our study had shown that infrared radiation could be effectively confined in 2D materials for photodetectors. The nano/micro-structures with graphene make promising materials for Si comparable photodetectors in midinfrared. The absorption of graphene in mid-infrared could be enhanced by using a light-trapping micro-holes structure and tuned to the desired region by adjusting the distance between the graphene layers. The resulting structure exhibits negative-index properties and has low losses. In addition to the resonance effect, the micro-holes at the surface reduce the reflection of the surface. Applications such as filters and polarizes also could use the combination of 2D materials and nano/micro-structures. The discussed structures could be used as tunable sensors or modulators changing the concentration of carriers in Si by pumping light and, thus, changing the structure's properties.

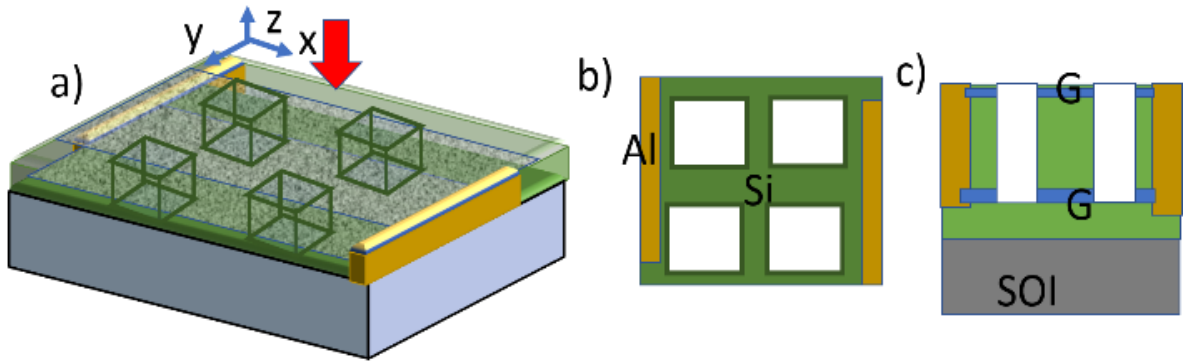

FIG. 1. The micro-hole structure with graphene layers (a) view from the top (b) view from the side

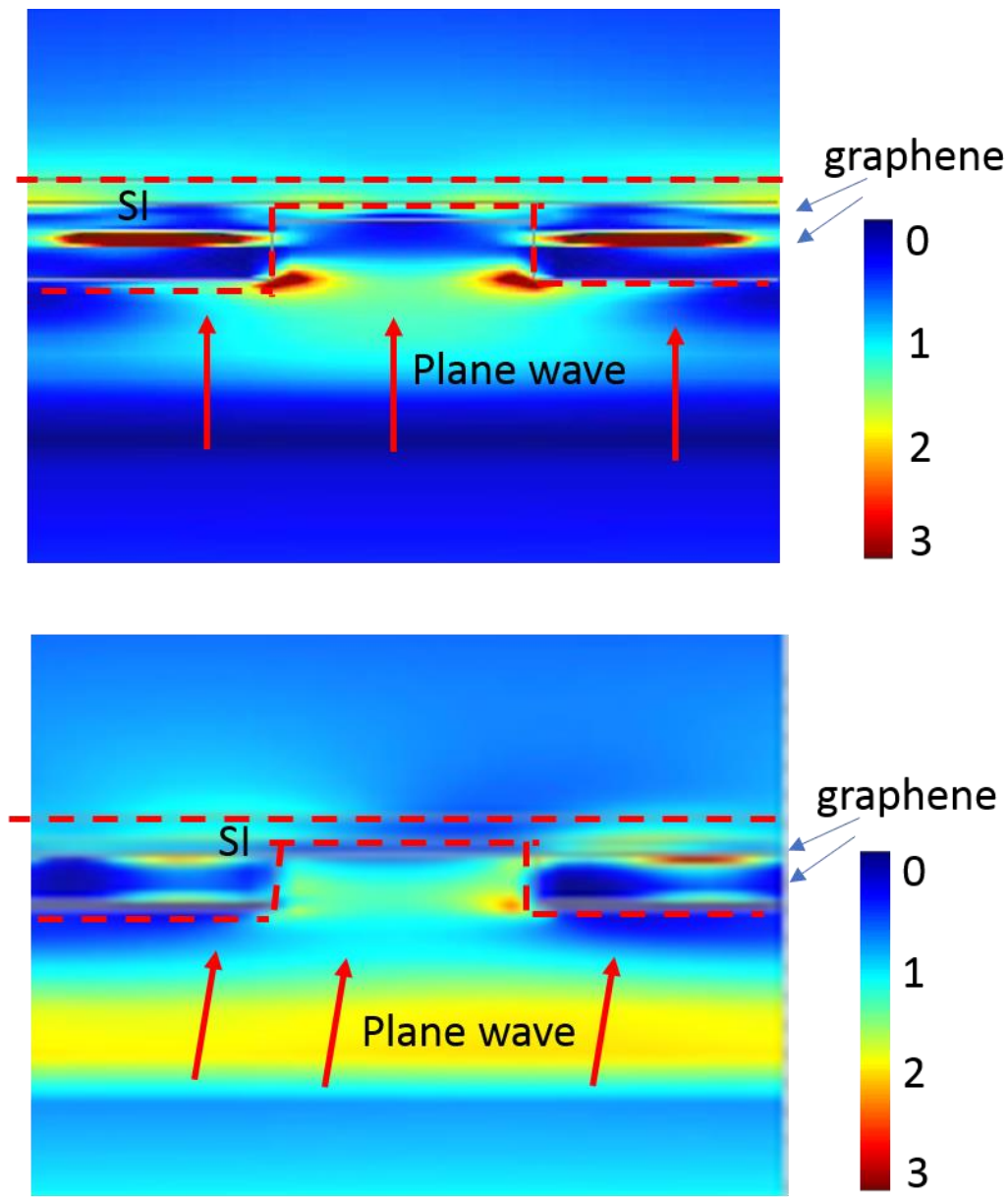

FIG. 2. Amplitude of the Poynting vector distribution shows plasmons at the graphene layers: top – normal incident illumination, bottom – 5degree incident illumination

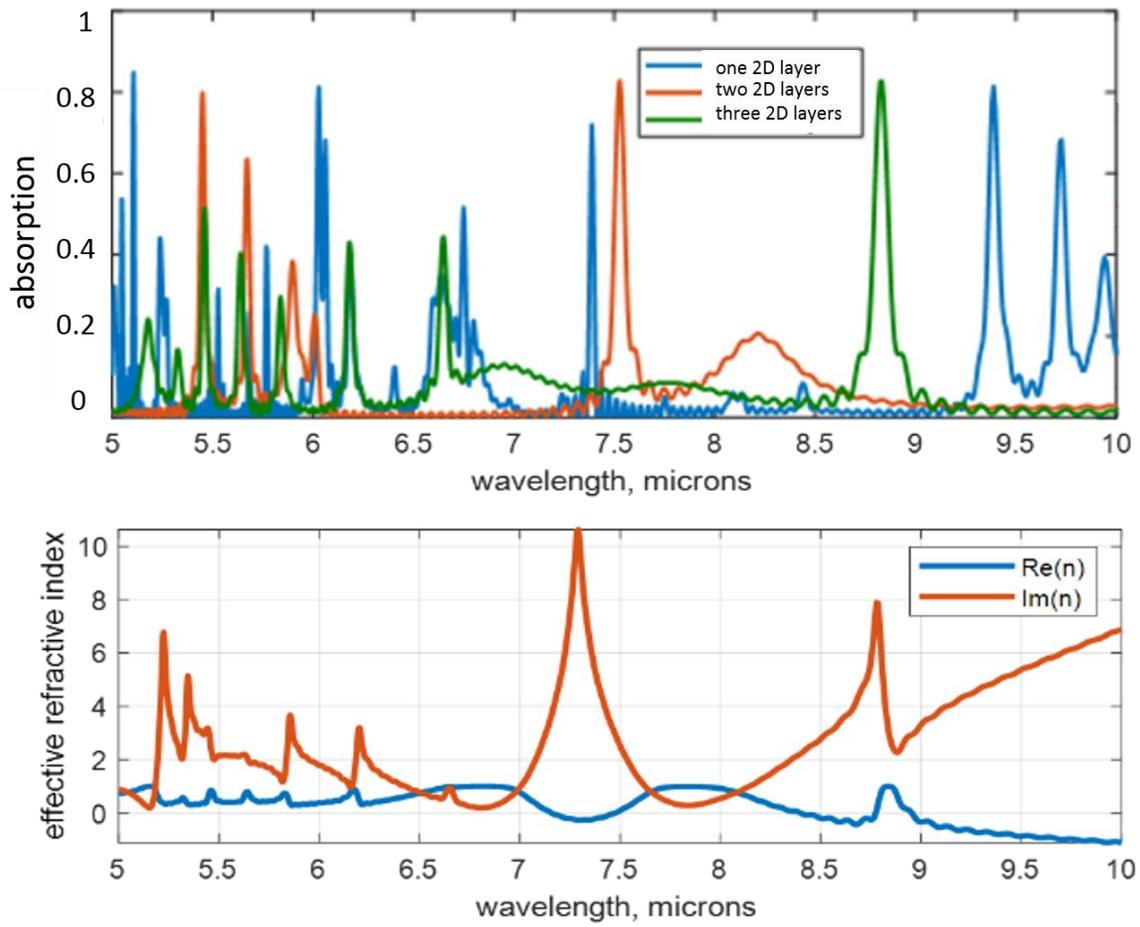

FIG. 3. Top: Absorption for 1 (blue), 2 (red), and 3 (green) graphene layers in Si; bottom: effective refractive index of the structure with 3 graphene layers

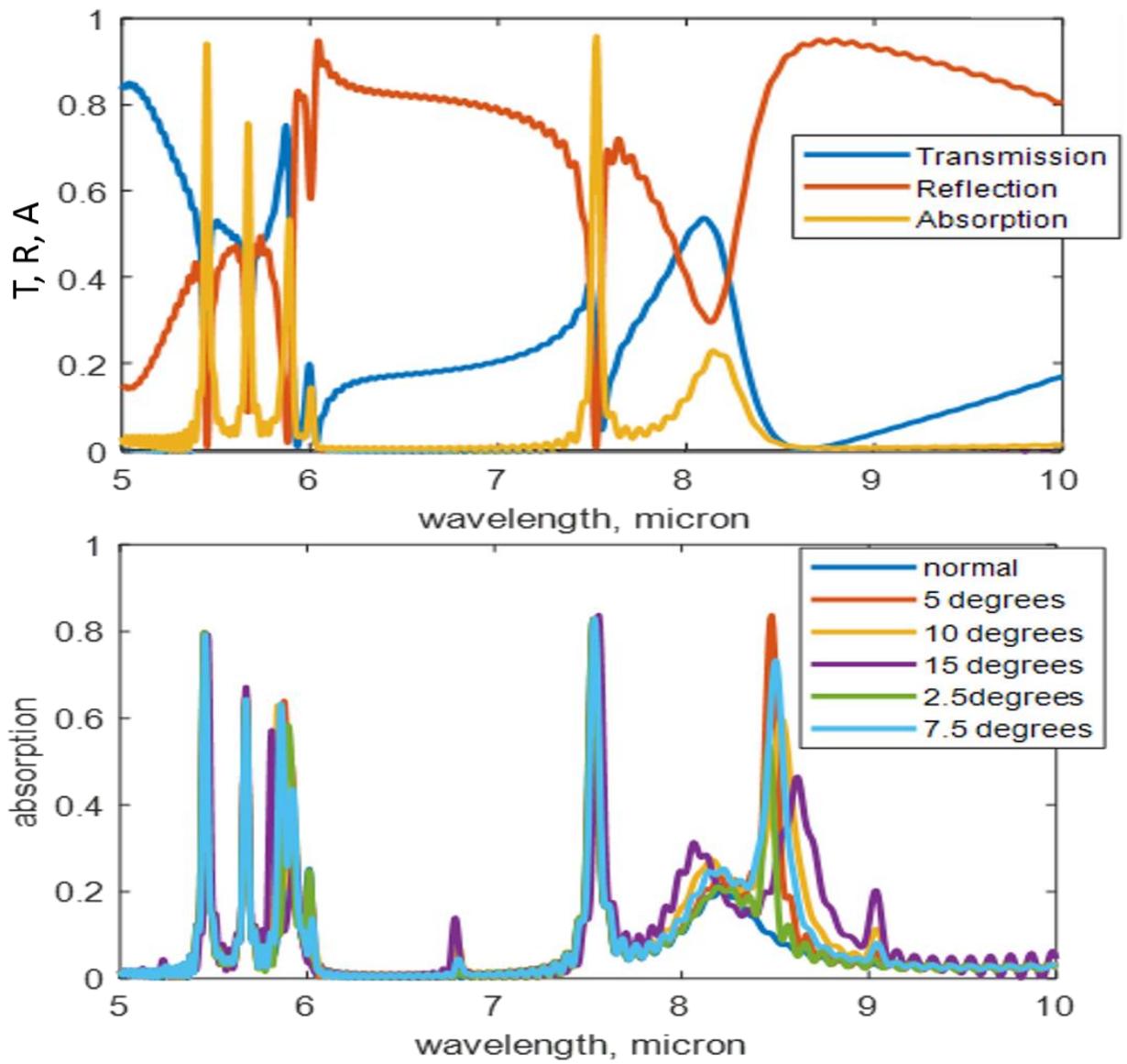

FIG. 4. FDTD simulations: top – Transmittance, reflectance, and absorption for the normal incidence, bottom – absorption for the different incident angles from 0 to 20 degrees.